\documentclass{rmf-d}
\usepackage{nopageno,rmfbib,multicol,times,epsf,amsmath,amssymb,cite}
\usepackage[latin1]{inputenc}
\usepackage[]{caption2}
\usepackage{graphics}

\usepackage {nopageno, rmfbib, multicol,times,epsf,amsmath,amssymb,cite}
\usepackage[latin1]{inputenc}
\usepackage[]{caption2}
\usepackage{graphics}
\usepackage[demo]{graphicx}
\usepackage{hyperref}
\usepackage{lipsum}
\usepackage{wrapfig, blindtext}
\usepackage{comment}
\usepackage{tabularx}

\usepackage{lipsum}
\usepackage{microtype}

\usepackage{amssymb}
\usepackage{amsmath}
\usepackage{graphicx}
\usepackage{dcolumn}
\usepackage{bm}
\usepackage{epsfig}
\usepackage{amssymb} 
\usepackage{color}
\usepackage{datetime}
\usepackage{wasysym}
\usepackage{slashed} 
\usepackage[mathscr]{euscript}
\usepackage{hyperref} 
\usepackage{tikz}

\newcommand{\bal}{\begin{align}}
\newcommand{\eal}{\end{align}}
\newcommand{\beq}{\begin{eqnarray}}
\newcommand{\eeq}{\end{eqnarray}}
\newcommand{\nneeq}{\nonumber \end{eqnarray}}

\newcommand{\nn}{\nonumber \\}
\newcommand{\es}{& = &}

\newcommand{\cM}{ {\cal M} }

\newcommand{\cL}{ {\cal L} }

\def\rmfcornisa{}

\def\rmfcintilla{{ } { } }
\clearpage \rmfcaptionstyle 
\begin{document}
\title{ An estimation of non-valence contributions to form factors of heavy-light mesons 
\vspace{-6pt}}
\author{ Mar\'ia G\'omez-Rocha }
\address{ Departamento de F\'isica At\'omica, Molecular y Nuclear
and Instituto Carlos I de F\'isica Te\'orica y Computacional
Universidad de Granada, E-18071 Granada, Spain   }
\author{Oliver Heger }
\address{ILF Consulting Engineers Austria GmbH, A-8074 Graz, Austria}
\author{ Wolfgang Schweiger}
\address{Institute of Physics, University of Graz, A-8010 Graz, Austria }
\author{ }
\address{ }
\author{ }
\address{ }
\author{ }
\address{ }
\maketitle

\begin{abstract}
\vspace{1em} 
We study the influence of non-valence quark-pair contributions in weak transition form factors of heavy-light mesons. Form factors are first calculated for spacelike momentum transfers in a reference frame where such contributions are suppressed. Analytic continuation to the timelike region and a comparison with the direct decay calculation, done with pure valence degrees of freedom, provides an estimate of the role that quark-pair contributions may play. We use the point form of relativistic quantum mechanics, which is particularly useful when treating heavy-light systems.
 \vspace{1em}
\end{abstract}
\keys{  Form factors, Few-body systems, Particle decays, Heavy-light mesons, Relativistic quantum mechanics   \vspace{-4pt}}

\begin{multicols}{2}

\section{Introduction}

The point-form~\cite{Dirac:1949cp} of relativistic quantum mechanics has been successfully employed to calculate hadron form factors within the framework of constituent quark models~\cite{Biernat:2007dn,Biernat:2014dea,Biernat:2011mp,Biernat:2010tp,Biernat:2009my,Gomez-Rocha:2014aoa,Heger:2021gxt,GomezRocha:2012zd,Gomez-Rocha:2013bga,Gomez-Rocha:2012wqk,Gomez-Rocha:2011jnc}. In this formalism the physical process in which a particular form factor is measuered is described in a Poincar\'e invariant way by means of the Bakamjian-Thomas construction~\cite{Bakamjian:1953kh}. In the point-form version of the Bakamjian-Thomas construction the (interacting) four-momentum operator $\hat P^\mu$ can be expressed as a product of an interaction-dependent mass operator and a free four-velocity operator,
\begin{eqnarray}
\hat P^\mu 
\es 
\hat \cM \, \hat V^\mu_{\text{free}}
\ = \ 
\left(\hat \cM_{\text{free}} + \hat \cM_{\text{int}}\right)\, \hat V^\mu_{\text{free}} \ ,
\label{eq:MV}
\end{eqnarray}
whereas the rest of the generators of the Poincar\'e algebra (rotations and boosts) remain free of interactions. This makes it particularly simple to add angular momenta and to boost wave functions, in contrast to other forms of relativistic dynamics.

With the velocity operator being free of interactions, the entire dynamics of the system is encoded in the mass operator $\hat \cM$. In order to account for a flavor change and the emission and absorption of gauge bosons in the description of electroweak processes, we adopt a coupled-channel framework in which the mass operator $\hat \cM$ acts on a direct sum of multiparticle Hilbert spaces. 
The diagonal matrix elements of $\hat \cM$ represent the kinetic energies of the elementary particles in the corresponding channel, while the non-diagonal entries are vertex operators $\hat K_{i\to j}$ and $\hat K_{j\to i} =\hat K^\dagger_{i\to j}  $ that account for the emission or absorption of gauge bosons and, therefore, the transition from one channel to another. 

The most convenient basis to represent operators within this framework consists of \textit{velocity-states} $|V ;\vec k_i , \mu_i \rangle$~\cite{Klink:1998zz}. These are multiparticle states at rest 
$|\vec k_i , \mu_i \rangle
\equiv
|\vec k_1 , \mu_1 ; \vec k_2 , \mu_2 ; ... ; \vec k_n , \mu_n \rangle$  \ ,
with $\sum_{i=1}^n \vec{k}_i = 0 $ and with $\mu_i$ being the $z$-projection of the (canonical) spin,which are boosted with the overall velocity $ V^\mu $ ($V^\mu V_\mu=1$) by means of a rotationless boost, i.e.
\begin{eqnarray}
|V ;\vec k_i , \mu_i \rangle:=
\hat U_{B_c(V)}|\vec k_i , \mu_i \rangle  \ .
\end{eqnarray}

Matrix elements of vertex operators are defined by means of the corresponding Lagrangian densities via~\cite{Klink:2000pp,Biernat:2009my,GomezRocha:2012zd}
\begin{eqnarray}
\langle V'; \vec k'_i , \mu'_i | \hat K |V; \vec k_i , \mu_i \rangle
&&\\ && \hspace{-1.5cm}= N V^0 \delta^3(\vec V - \vec V')
\langle \vec k'_i , \mu'_i | \hat \cL_{\text{int}} | \vec k_i , \mu_i \rangle \, , \nonumber
\label{eq:vertex}
\end{eqnarray}
where the factor $N$ is a normalization factor determined by the normalization of the velocity states. The delta function guarantees the conservation of the overall velocity of the system at every interaction vertex, which is demanded by Eq.~(\ref{eq:MV}). 

It has been argued~\cite{Keister:1992wq} and shown~\cite{Gomez-Rocha:2012wqk,Gomez-Rocha:2013bga} that the point form of relativistic dynamics is particularly suitable for the description of heavy-light systems. Hadronic systems with a constituent much heavier than the other(s) exhibit an additional symmetry known as heavy-quark symmetry~\cite{Neubert:1993mb}. In such systems the velocity of the heavy-light bound state is approximately conserved and it is thus no longer a dynamical degree of freedom~\cite{Neubert:1991xw,Neubert:1993mb}.

The present work addresses the description of form factors of heavy-light mesons  within the context of constituent quark models. In particular, we want to study the influence of a $(Q\bar{q})(q^\prime\bar{q}^\prime)$  non-valence Fock component on weak decay form factors. A  non-valence $(Q\bar{q})(q^\prime\bar{q}^\prime)$ Fock component in the decaying meson may give rise to a vector-meson-dominance-like decay mechanism in which the $W$ does not directly couple to the heavy quark of the valence Fock state, but the non-valence Fock state rather splits into the valence component of the final meson and an intermediate vector meson which subsequently is converted into the $W$ boson. This kind of contribution, often termed as \lq\lq Z-graph\rq\rq , is by no means negligible as compared to the pure valence contribution, in particular if one approaches the zero-recoil point. Semileptonic weak decays provide information on meson transition form factors for timelike momentum transfers, with neutrino-meson scattering one rather explores the spacelike momentum-transfer region. For spacelike momentum transfers the Z-graph contribution plays a minor role as compared to timelike momentum transfers. One can even find a reference frame, the so-called \textit{infinite-momentum frame} (IMF), in which it is suppressed and the valence Fock-state provides already a complete description of the form factors. Similarly, the Z-graph contribution is also suppressed in the heavy-quark limit, in which the masses of the heavy quarks go to infinity and heavy-quark symmetry is restored. In the front form of relativistic dynamics, the Z-graph contribution is eliminated by choosing a $q^+=0$ Drell-Yan-West frame. The IMF is a particular example for such a frame. For timelike momentum transfers, however, one always has $q^+>0$ and thus it is not possible to exploit the advantages of the $q^+=0$ Drell-Yan-West frame or of the IMF and one has to be concerned about the Z-graph contribution.

Since an explicit calculation of the Z-graph contribution would require additional modeling, we rather follow another strategy to estimate its size in the timelike momentum-transfer region:

\noindent 1) Extract analytic expressions for the meson transition form factors from the neutrino-meson scattering amplitude in the IMF, where the valence contribution is supposed to provide already a fairly complete description.

\noindent 2) Continue these form factor expressions analytically to timelike momentum transfers. Provided that the analytic continuation is done correctly, it should also provide a complete description of the decay form factors. 

\noindent 3) Compare the numerical results from analytic continuation with those from the direct decay calculation done within the pure valence-quark picture to estimate the possible role of non-valence contributions.


\section{Weak $B\to D$ transition form factors}
%
We illustrate our procedure by means of the $B\to D$ transition. For a more comprehensive study, the interested reader may consult Ref.~\cite{Heger:2021gxt}.

\subsection{Spacelike momentum transfer}
\label{spacelike}

First of all, we study $B\to D$ transition form factors for space-like momentum transfers, $Q^2=-q^2<0$, as can be measured in  neutrino-meson scattering, $\nu_e B^- \to e^- D^0$. This reaction involves the exchange of a $W$ boson. Taking into account all states that occur in the course of the scattering process, i.e. $|\nu_e, b,\bar u \rangle$, $|e, W^+,b, \bar u \rangle$, $|e, c,\bar u \rangle$ and $|\nu_e, W^-, c,\bar u \rangle$, one ends up with a four-channel eigenvalue problem for the mass operator. The diagonal matrix elements of the mass operator contain the kinetic energies of the particles and, in addition, an instantaneous confining force between the quark and the antiquark. Eliminating the two channels containing a $W$ by means of a Feshbach reduction, on ends up with an optical $1W$-exchange potential that describes the transition from the $\nu_e b\bar u$ to the $e c \bar u$ channel. Taking matrix elements of this optical potential between the incoming $|\nu_e B^-\rangle$ and the outgoing $|e^- D^0\rangle$ states, gives already the $\nu_e B^- \to e^- D^0$ scattering amplitude in leading order of the weak coupling. The weak hadron transition current can then easily be extracted from this scattering amplitude in a unique way. It is an overlap integral containing the $Q\bar{q}$ wave functions of the $B$ and the $D$,  the weak quark current and some kinematical factors. The wave functions contain also a Wigner-rotation factor as a result of  boosting the mesons at rest to their respective momenta~\cite{Heger:2021gxt}.  Knowing the expression for the hadronic current as a function of the incoming and outgoing momenta $J^\mu(p_D,p_B)$, the form factors can be extracted from a covariant decomposition of the current:
\begin{eqnarray}
\tilde{J}^{\nu}_{B\rightarrow D}(p_{D}; p_{B}) &&
\nn
&&
\hspace{-1.7cm}= \left( (p_{B}+p_{D})^{\nu} - \frac{m_{B}^2-m_{D}^2}{q^2}q^{\nu}\right) F_1(q^2,s)
\nonumber\\
&&\hspace{-1.7cm}\phantom{=} + 
\frac{m_{B}^2-m_{D}^2}{q^2}q^{\nu}F_0(q^2,s)\, .
 \label{eq:Jcovdec}
\end{eqnarray}
At this point it is necessary to mention that the Bakamjian-Thomas construction, in particular the conservation of the overall velocity in Eq.~(\ref{eq:vertex}, gives rise to wrong cluster properties. This drawback is common to all forms of dynamics, and not a particular problem of the point form~\cite{Keister:1991sb}. As a consequence, the gauge-boson-hadron vertices do not only depend on the incoming and outgoing boson momenta, but also on the lepton momenta. The form factors are thus not only functions of the 4-momentum transfer squared, but they also exhibit a dependence on Mandelstam $s$, the invariant mass squared of the neutrino-meson system. This fact can be also interpreted as a frame dependence of the $W B\to D$ subprocess. One may then consider two particularly interesting frames. Taking the minimum value of $s$ necessary to reach a particular $q^2<0$, one ends up with the so-called Breit frame (BF). The other frame, corresponding to $s\to \infty$, is the IMF mentioned already, . 

In order to perform numerical calculations we use a harmonic-oscillator quark-antiquark wave function and adopt the parameters of an analogous calculation performed within the front form of relativistic dynamics~\cite{Cheng:1997}:
\begin{eqnarray}
\psi_\alpha (\kappa)
\es
{2 \over \pi^{4\over 2} a_\alpha^{3\over 2}} e^{- {\kappa^2 \over 2 a_\alpha^2}} \ 
\end{eqnarray}
with $a_B=0.55$ for the $B$ meson and $a_D=0.46$ for the $D$ meson.

An extended numerical study performed in~\cite{Heger:2021gxt} shows small differences between the IMF and the BF predictions for $B\to D$ transition form factors. Numerical results obtained for other reactions indicate that the frame dependence decreases if the size of the three-momentum of the final-state meson increases.  In our approach there are two obvious sources for this frame dependence: on the one hand wrong cluster properties, on the other hand a missing Z-graph contribution in frames different form the IMF. Our studies, like foregoing work~\cite{Gomez-Rocha:2012wqk,Biernat:2009my,Biernat:2014dea}, suggest that the frame dependence of the form factors goes away with increasing value of the Mandelstam $s$. The form factors extracted in the IMF therefore seem to provide the most realistic picture of the electroweak meson structure for spacelike momentum transfers as long as one sticks to a pure valence-quark description.

\subsection{Timelike momentum transfer}

Weak transition form factors for timelike momentum transfers, $q^2\geq 0$, are experimentally measured in semileptonic weak decay processes, e.g. for the $B\rightarrow D$ transition in the $B^-\rightarrow D^0 e\, \bar{\nu}_e$ decay. Since scattering amplitudes are functions of Mandelstam $s$ and $t=q^2$, it is possible to continue analytically from $q^2\le 0$ to $q^2\ge 0$ in the expressions for the currents and the form factors. Provided that the analytic continuation is done correctly, analytic continuation of the form factors extracted in the IMF should already give physically sensible form factor results for time-like momentum transfers. In order to estimate the size of the Z-graph contribution and possible cluster-separability violating effects -- both become negligible in the IMF -- for time-like momentum transfers, we can then compare with the outcome of a direct decay calculation done within the pure valence-quark picture. 

It is straightforward to describe decays in our coupled-channel formalism. One has to define a mass operator which acts on $|\bar u b\rangle$, $|\bar u c W\rangle$, $|\bar u b W e \bar \nu_e\rangle$, $|\bar u c e \bar \nu_e\rangle$ states and calculate the decay amplitude in leading order of the weak coupling (in analogy to the scattering amplitude). Again, the decay current can be extracted from the decay amplitude in a unique way and from the decay current the form factors are identified using the covariant decomposition (\ref{eq:Jcovdec}). Form factors calculated for timelike momentum transfers in this way are, however, just functions of the momentum transfer squared; there is no $s$-dependence in this case, since the invariant mass of the decaying system is the mass $m_B$ of the decaying meson, which is fixed. There is no freedom in this case for choosing the decay kinematics, apart of rotations or boosts, which, however, do not affect the covariant decomposition ~(\ref{eq:Jcovdec}).

\section{Numerical analysis}

\begin{figure}[t!]
\centering
\includegraphics[scale=0.22]{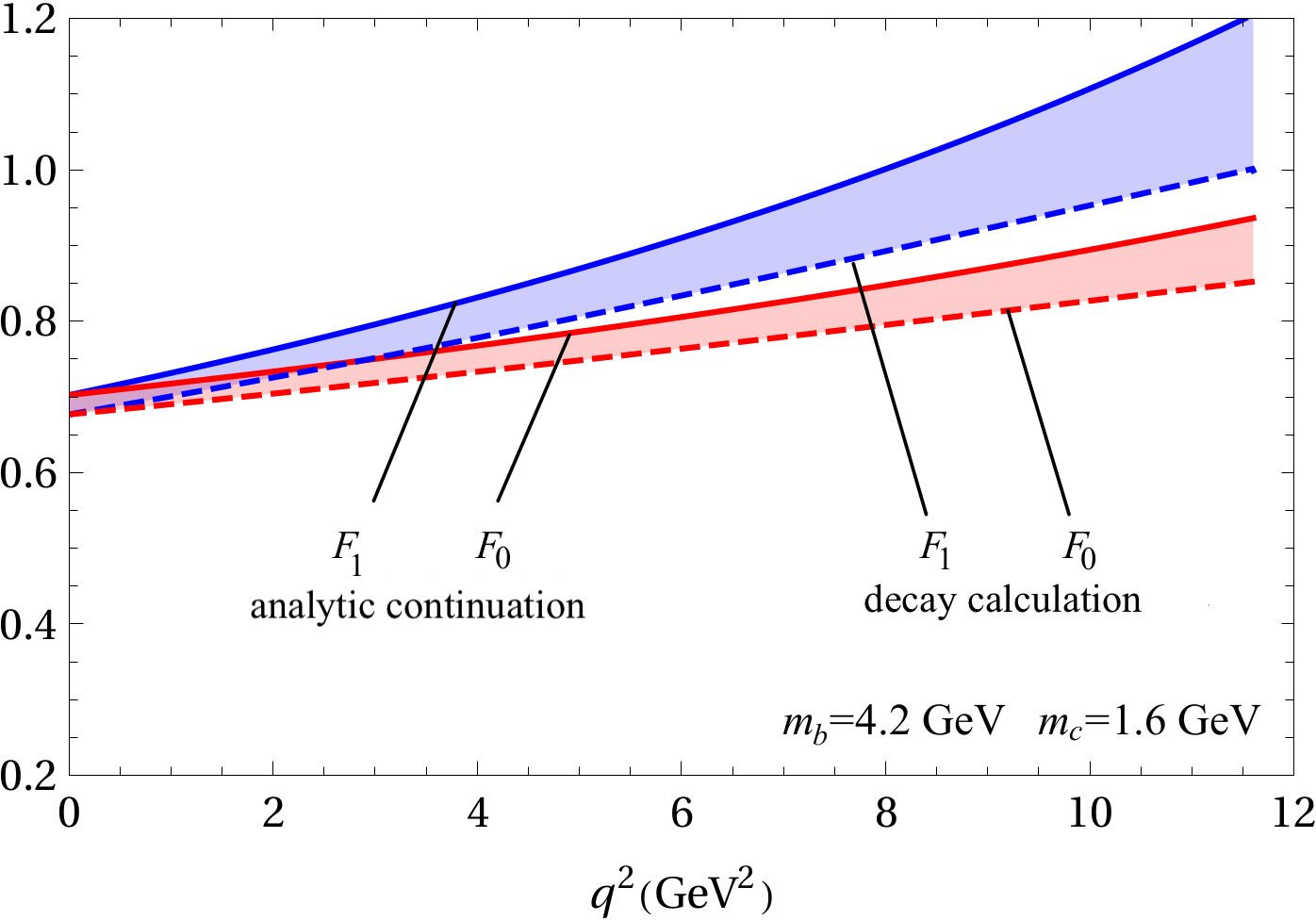}\hspace{1.5cm}
\includegraphics[scale=0.225]{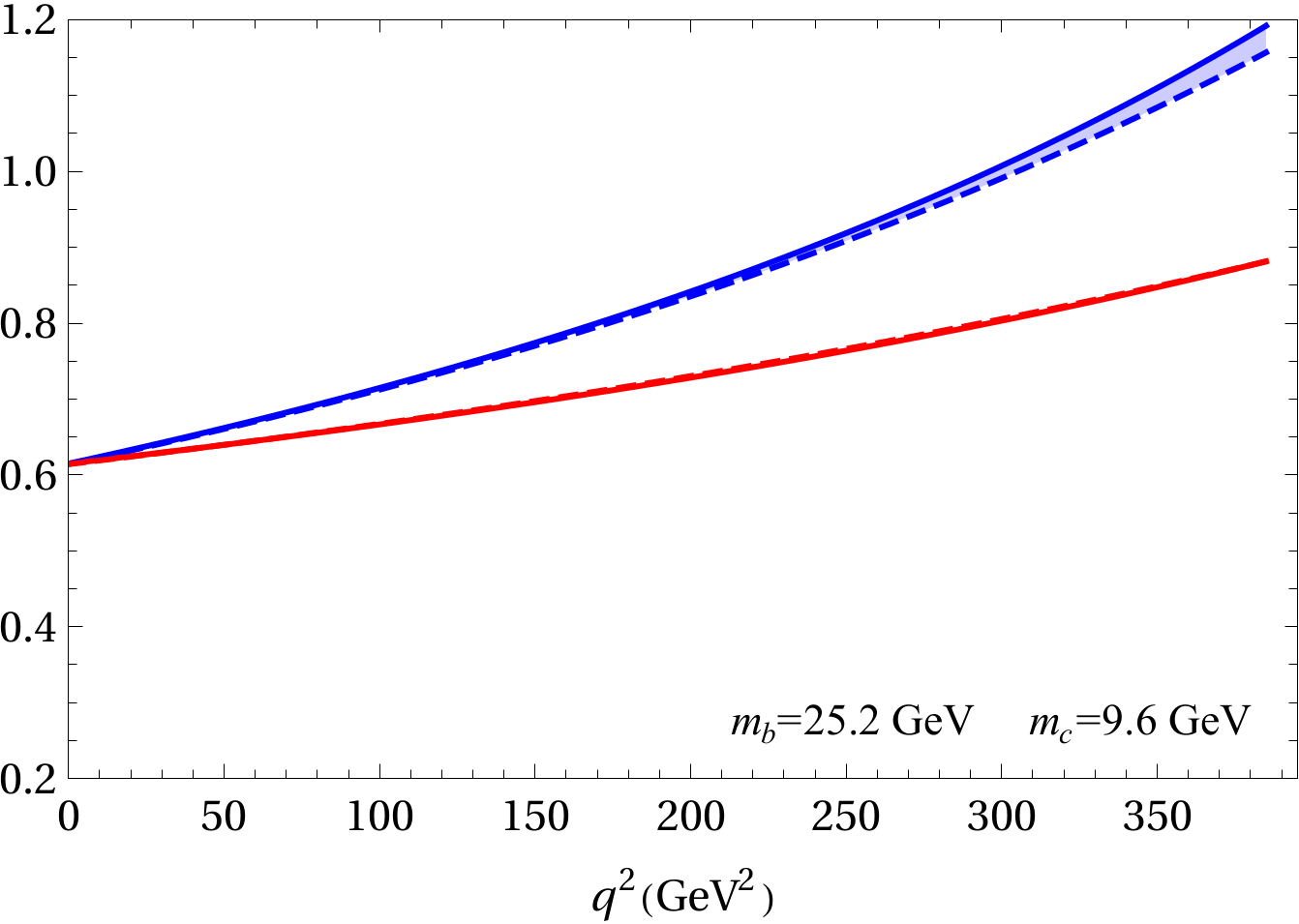}
\caption{Weak $B\rightarrow D$ transition form factors $F_0$ and $F_1$ as resulting from analytic continuation of the IMF result (solid lines) and from the direct decay calculation (dashed lines). Left figure: Physical meson and quark masses. Right figure: Heavy quark masses are multiplied by a factor 6 and meson masses are taken to be equal to the heavy quark masses. The shaded area indicates the order of magnitude of the Z-graph contribution, which is missing in the direct decay calculation.}
\label{fig:comparison}
\end{figure}

Predictions for the weak $B\rightarrow D$ transition form factors $F_0$ and $F_1$, as resulting from analytic continuation of the IMF calculation and the decay calculation, are compared in Figure~\ref{fig:comparison}. Since the Z-graph and cluster-separabilty-violating effects vanish in the IMF, the analytically continued IMF results are supposed to provide a more complete description of the timelike form factors than the direct decay calculation. The differences therefore give us an estimate of the size of such effects in the timelike momentum transfer region.  They become larger with increasing $q^2$.

Whereas the left panel shows results for physical quark and meson masses, the right panel exhibits predictions for upscaled masses of the heavy quarks and the mesons. This allows us to check, whether the form factors exhibit the expected behavior in the heavy-quark limit.
One knows from heavy-quark symmetry arguments~\cite{Neubert:1993mb,Gomez-Rocha:2012wqk} that Z-graphs involving heavy quarks should be suppressed in the heavy quark limit. This is reflected in our results and illustrated in the right panel of Figure~\ref{fig:comparison}. When the constituent quark masses are multiplied by a factor 6 and the meson masses are set equal to the heavy quark masses, the differences attributed to the Z-graph contribution and possible cluster-separability violating effects tend to vanish for both form factors $F_0$ and $F_1$.

In order to provide some numerical comparison with experiment, we consider the slope of $F_1$ as a function of the product of four-velocities $v_B\cdot v_D$ at zero recoil:
\begin{eqnarray}
\rho_D^2 := - {F_1'(v_D \cdot v_B=1) \over F_1(v_D \cdot v_B=1)} \ ,
\end{eqnarray}
where $F_1'$ means differentiation with respect to $v_D \cdot v_B$. 
The experimental value for this quantity, $\rho^2_{D\,\text{exp}} = 1.131 \pm 0.033 \ $,  provided by the heavy-flavor averaging group~\cite{HFLAV:2019otj},  is very close to our result obtained in the IMF by analytic continuation: 
$
\rho^2_{D\,\text{IMF}} = 1.07 \ .
$
It is worth noticing that this result differs considerably from the result obtained in the direct decay calculation, which involves only valence degrees of freedom, $\rho_{D\, {\text{direct}}}^2 = 0.55$. This supports our argument that the physically most complete description within a pure valence-quark picture is achieved in the IMF. It remains to be checked whether the direct decay calculation can be reconciled with the IMF result by adding the Z-graph contribution.  

\subsection{Meson pole}

\begin{figure}[h]
\vspace{-0.3cm}
\begin{minipage}[c]{0.55\textwidth}\begin{center}
\includegraphics[scale=0.4]{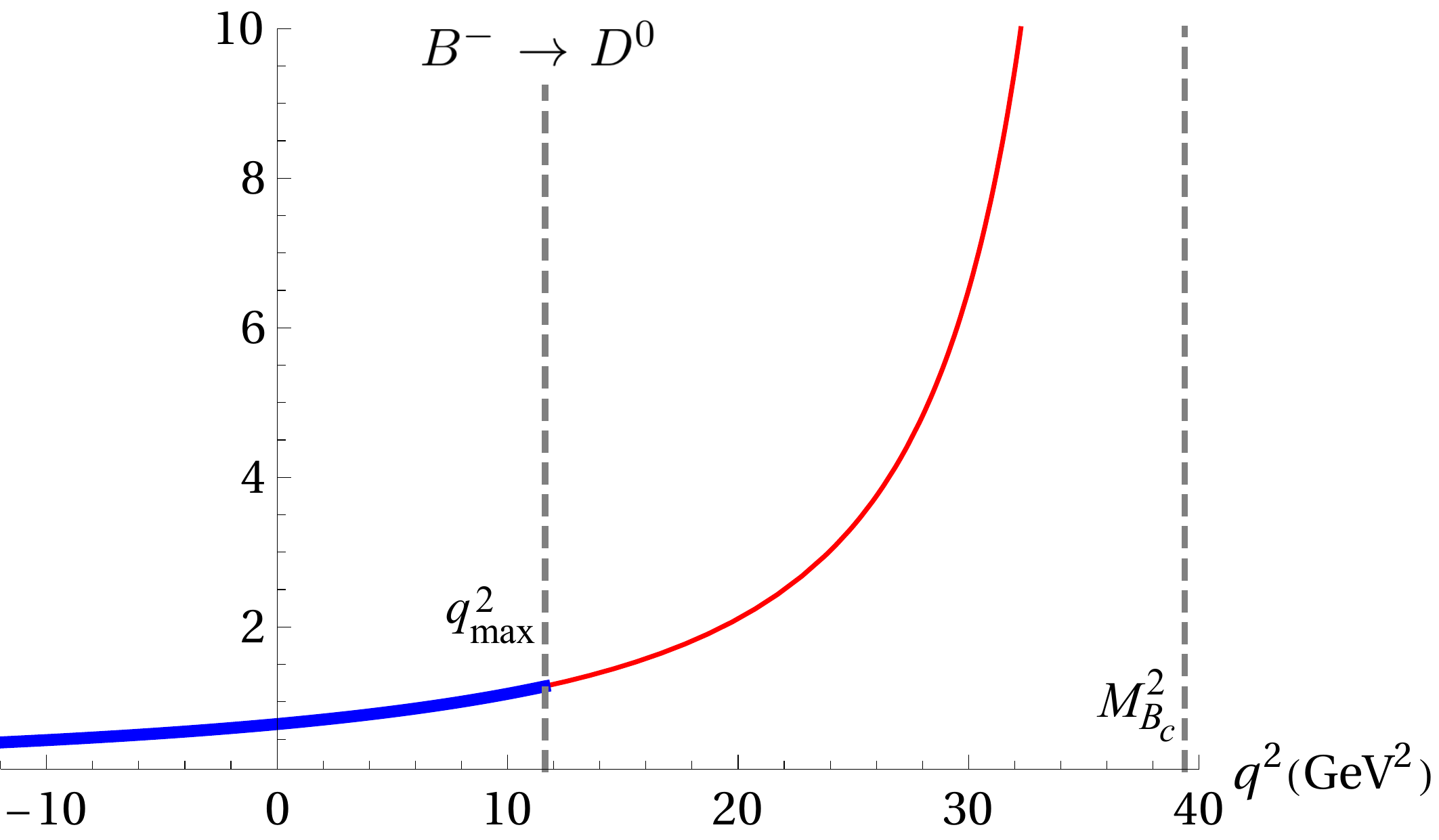}
\end{center}\end{minipage}
 \hfill
\begin{minipage}[l]{0.4\textwidth}
\caption{
The thick, blue line shows the transition form factor $F_1$ for space- and timelike momentum transfers obtained by analytic continuation of the infinite-momentum-frame result. The red thin line is the pole fit, Eq.~(\ref{eq:pole}) with $\alpha=1.55$ and $m_{\text{pole}}=m_{B^*_c}\approx m_{B_c}=$6.274 GeV. The dashed lines indicate the position of $q^2_{\text{max}}$ and $m^2_{\text{pole}}$.  } 
   \label{fig:resonance}
\end{minipage}
\end{figure}

A more extended record of results including decay processes different from $B\rightarrow D$ is given in Ref.~\cite{Heger:2021gxt}.
This study showed that the difference between the IMF result and the direct decay calculation becomes even larger when the mass of the final-state meson becomes smaller. For instance, for $B\to \pi$ and $D\to \pi$ transitions a pronounced pole-like behavior is observed for the IMF result near the zero recoil point, $q^2_{\text{max}}=(m_{B(D)}-m_\pi)^2$. Within our constituent-quark model such a behavior can be understood, if one allows for a non-valence $Q\bar{q} q ^\prime \bar{q}^\prime$ component in the decaying meson. Taking into account confinemnet, such a component can decay into the final state $q ^\prime\bar{q}$ meson and a $Q\bar{q}^\prime$ vector meson which fluctuates into the $W$ boson giving rise to a vector-meson-dominance like decay mechanism. Such a decay mechanism was also considered in Ref.~\cite{Cheung:1996qt}. The authors analyzed the direct decay (involving the valence Fock state)  and the vector-meson-dominance like decay mechanism separately for the semileptonic $B\to \pi$ decay. They found that the sum of both contributions is well approximated in the whole range $0\leq q^2\leq q^2_{\mathrm{max}}$ by a function of the form
\begin{eqnarray}
F^{\text{pole}}_1 (q^2)
\es
{F_1(0) \over \left( 1 - {q^2\over m^2_{\text{pole}}} \right)^\alpha}  \ ,
\label{eq:pole}
\end{eqnarray}
with $\alpha = 2.0$ and $m_{\text{pole}}=5.6-5.8$ GeV, depending on the strength of the $B^\ast B\pi$ coupling.
They observed that the valence contribution to $F_1$ dominates at $q^2=0$, wheras the non-valence $B^*$ pole contribution was dominant near the zero-recoil point $q^2_{\mathrm{max}}$. Following Ref.~\cite{Cheung:1996qt} and using our result at zero momentum transfer $F_1(0)=0.70$ we have parameterized our IMF results employing Eq.~(\ref{eq:pole}) . But we fixed  the pole mass to the lightest vector-meson mass and just left the power $\alpha$ as a free parameter. The outcome of this parametrization is shown in Figure~\ref{fig:resonance} for the $B\rightarrow D$ decay. We find that the parameterization~(\ref{eq:pole})  with $m_{\text{pole}}=m_{B_c^\ast}\approx m_{B_c}=6.274$~GeV and $\alpha=1.55$ reproduces the analytically continued IMF result surprisingly well. 
This holds also for other decay processes~\cite{Heger:2021gxt}. The form factor tends to exhibit a monopole-like behavior, the closer the pole is to the zero recoil point. 

\section{Conclusions}


It is very remarkable that the analytic continuation of the IMF results provides a behavior of the timelike meson-transition form factors that resembles a vector-meson-dominance like decay mechanism, although we started with a pure valence picture and did not include any additional dynamical input. How this can happen, whether the discrepancy between the analytically continued IMF result and the direct decay calculation can be explained by including non-valence degrees of freedom in the direct decay calculation and whether cluster-separability-violating effects play a minor role, as sometimes asserted~\cite{Keister:2011ie}, will be the subject of further investigations.

\section*{Acknowledgments}
M.G.R. acknowledges supported received from the Spanish MINECO's Juan de la Cierva-Incorporaci\'on programme, Grant Agreement No. IJCI-2017-31531, Junta de Andaluc\'ia FQM-225, and project PID2020-114767GB-I00 funded by MCIN/AEI/10.13039/501100011033.

\end{multicols}

\begin{multicols}{2}

\end{multicols}

\vfill\break

\begin{multicols}{2}

\medline

\bibliography{PF}{}

\bibliographystyle{unsrt}

\end{multicols}
\end{document}